\documentclass[aps,preprint]{revtex4-2}

\usepackage[english]{babel}
\usepackage[utf8x]{inputenc}
\usepackage[T1]{fontenc}

\usepackage[a4paper,top=3cm,bottom=2cm,left=3cm,right=3cm,marginparwidth=2cm]{geometry}

\usepackage{amsmath}
\usepackage{graphicx, subfigure}
\usepackage[colorinlistoftodos]{todonotes}
\usepackage[colorlinks=true, allcolors=blue]{hyperref}
\usepackage{float}
\usepackage{comment}

\begin{document}

\title{A Mathematical Model to Capture Urbanization Trajectory\\ Induced by Economic Inequality}

\author{Neeraj Pandey}
\affiliation{Department of Physics, School of Advanced Engineering,\\ UPES, Dehradun 248 007, India}

\author{Abhineet Agarwal}
\affiliation{Centre for Theoretical Physics,\\ Jamia Millia Islamia New Delhi-110025, India}

\author{Raju Roychowdhury}
\affiliation{Department of Physics and Astrophysics, University of Delhi, Delhi 110 007, India}

\author{Karmeshu}
\affiliation{Centre for Stochastic Modelling and Simulation,\\ UPES, Dehradun 248 007, India,}
\affiliation{Shiv Nadar University, Greater Noida, UP, 201310, India}

\author{Parth Pratim Pandey}
\email{parth.pratimji@gmail.com}
\affiliation{Department of Physics, School of Advanced Engineering,}
\affiliation{Centre for Stochastic Modelling and Simulation,\\ UPES, Dehradun 248 007, India}

\date{\today}

\begin{abstract}
Analysis of the urban population fraction data for sixteen populous countries over the last fifty years reveals a universal increase in urbanization - exhibiting four qualitatively distinct temporal patterns: (i) continuously accelerating growth, (ii) continuously decelerating growth, (iii) two-phase growth transitioning from acceleration to deceleration, and (iv) two-phase growth transitioning from deceleration to acceleration. To understand the origin of these diverse urbanization trajectories, we develop a simple coarse-grained model in which a country is segregated into two regions, a rural and an urban region. Population in each region evolves due to natural (sexual) growth and migration from rural to urban areas, with the migration rate governed by economic inequality, quantified through the difference in GDP per capita between the two regions. The GDP per capita of both regions is assumed to grow exponentially with distinct rates. We demonstrate that this minimal model, involving four dynamical variables and a small number of demographic and economic parameters, is capable of reproducing all four empirically observed urbanization patterns. Assuming demographic and economic parameters remain approximately constant over a 50-year timescale, we estimate coarse-grained parameters for the United States using empirical data and obtain optimized values that accurately reproduce its observed urbanization trajectory. Our results highlight how simple demographic–economic interactions can generate rich and diverse urbanization dynamics.
\end{abstract}

\maketitle

\section*{Introduction}
Rural to urban migration has been steadily increasing in most countries over the years. Thus, it was not unexpected when, between 2007 and 2009, humanity experienced a historic shift in global settlement patterns - the world's urban population surpassed the rural population for the first time \cite{unfpa2007unleashing}. Although urbanization is associated with improved economic prospects and living standards, over-urbanization can place significant stress on the environment, compromise sustainability, and deepen social disparities. Thus, understanding the key factors driving the urbanization dynamics is key to developing efficient and effective strategies to manage urban growth. Since the urban fraction is a fractional variable, it will always be bounded between 0 and 1, but being a complex variable, its trajectory will be influenced by a vast number of country-specific social, economic, demographic, political and environmental parameters. Furthermore, the socioeconomic landscape of a country changes with time, making these parameters dynamic.

Theoretical and mathematical models of rural–urban migration have been developed at multiple levels of abstraction to capture different mechanisms driving population flows. Early dual-sector models, most notably the Lewis and Harris–Todaro frameworks \cite{lewis2016economic, todaro1969model, harris1970migration}, explain migration as a rational response to expected income differentials between rural and urban regions, accounting for urban unemployment and equilibrium conditions. Gravity and spatial-interaction models take a more aggregate perspective, describing migration flows as proportional to population “masses” of origin and destination and inversely related to distance or migration costs \cite{tinbergen1962shaping, wilson2013entropy, greenwood1985human}.
At a more aggregated demographic level, urbanization has been modeled using replacement dynamics, in which coupled rural–urban population growth and transfer processes generate robust and empirically validated urbanization trajectories, capable of explaining cross-country variation and the persistence of urban growth over time \cite{karmeshu1988demographic, rao1989dynamics}. More recently, agent-based models explicitly simulate individual or household decision-making, social learning, and network effects, often reproducing Harris–Todaro–type equilibria as emergent outcomes \cite{espindola2006harris}. Complementary approaches include network models, which emphasize the role of social ties and information diffusion \cite{massey1993theories}, and continuum or reaction–diffusion models, which treat population densities and migration fluxes in a spatially continuous mathematical framework \cite{okubo2002diffusion}. Together, these approaches provide a spectrum of models linking micro-level behavior to macro-level rural–urban migration patterns.

\begin{figure}[h]
\centering
\includegraphics[width=14.5cm,height=12cm]{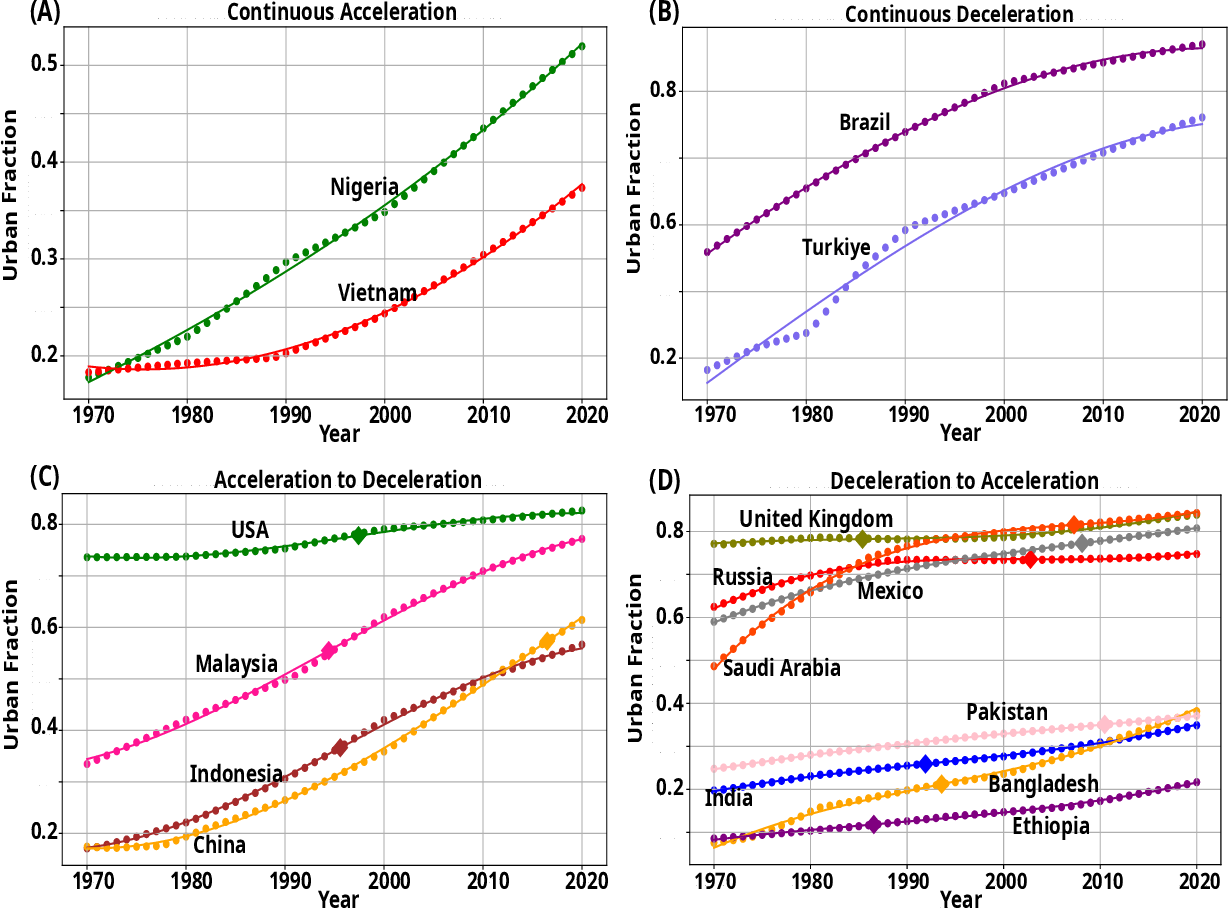}
\caption{{\bf Trend-wise separation of countries according to urbanization trajectory.} We (cubic spline) fit the urbanization data for 16 countries for the past 50 years (1970-2020) and segregated them into four categories. (A) \textit{Continuous Acceleration:} Countries like Nigeria and Vietnam displayed an accelerating rate of urbanization throughout this period. (B) \textit{Continuous Deceleration:} Brazil and Turkiye showed a decelerated increase in the urbanization curve. (C) \textit{Acceleration to Deceleration:} The data for USA, Malaysia, Indonesia, and China exhibited an accelerating increase initially, followed by a decelerated growth regime beyond an inflection point. (D) \textit{Deceleration to Acceleration:} On the other hand, the urban fraction of the United Kingdom, Russia, Mexico, Saudi Arabia, Pakistan, India, Bangladesh and Ethiopia initially increased with a decelerated rate but soon transitioned into an accelerated growth. The inflection points in categories (C) and (D) are highlighted on the curve (large diamond point).}
\label{fig:EmpiricalUrbanFrac}
\end{figure}

In this manuscript we first analyze the urban fraction data (the fraction of total population living in urban regions) for the last 50 years (1970 - 2020) for several of the world's most populous countries. This reveals four characteristically different trends (Fig. \ref{fig:EmpiricalUrbanFrac}) :- (1) \textit{Continuous Acceleration}: nations such as Nigeria and Vietnam experienced an accelerated increase in urbanization throughout this period; (2) \textit{Continuous Deceleration}: Brazil and Turkiye, on the other hand, displayed a decelerated growth of urbanization during this period; (3) \textit{Acceleration to Deceleration}: countries like Indonesia, Malaysia, USA and China first showed an accelerated increase of the urbanization curve, followed by a decelerated growth regime; (4) \textit{Deceleration to Acceleration}: lastly, India, Bangladesh, Ethiopia, Russia, Pakistan, United Kingdom, Mexico and Saudi Arabia showed the opposite trend, i.e., an initial phase of decelerated increase of the urban fraction followed by an accelerated regime. The above classification indicates that in the last 50 years, while an increasing urban fraction has been a common trend, the specific trajectory of increase varies for different countries, highlighting the diverse socio-economic and modernization pathways.

Being a complex phenomenon, the rate of rural-to-urban migration must be influenced by a wide range of factors. Some of the key factors are; improved access to education, healthcare, infrastructure, and public services \cite{greenwood1985human, desa2014world}. Typically, regions with higher GDP per capita tend to offer more developed public services and improved quality of life \cite{baldacci2004social,haryanto2021relationship,brueckner2021infrastructure,Singh_Cheemalapati_RamiReddy_Kurian_Muzumdar_Muley_2025}. Thus, for a minimal coarse-grained model, it can be argued that the Gross Domestic Product (GDP) per capita serves as an effective surrogate indicator for these variables. In this manuscript we construct a simple model for rural to urban migration, assuming that the difference in GDP per capita between two regions indicates the disparity in the availability of the above-mentioned basic services. Assuming this, we build a coarse-grained toy model for migration consisting of just two regions, a rural and an urban region, and show that the above mentioned urbanization trends emerge in this simple model depending upon the interplay of coarse-grained economic, population growth and migration parameters.

\section*{Results}
\subsection*{The Model}

We now describe our coarse-grained model to quantitatively capture the evolution of the urban population fraction over time. Consider two regions, a rural and an urban region, with population $N_r(t)$ and $N_u(t)$ at time $t$ respectively, following the rate equations:

\begin{subequations} \label{eq:AbsoluteRateEqns}
    \begin{align}
        \dot {N_{r}} &= \beta_{r}N_r(t) - kN_r(t)\biggl[ E_u(t) - E_r(t)\biggr], \label{eq:RuralRateEqns}\\
        \dot {N_{u}} &= \beta_{u}N_u(t) + kN_r(t)\biggl[ E_u(t) - E_r(t)\biggr]. \label{eq:UrbanPopEqn}
    \end{align}
\end{subequations}

The first term in the two equations represents an increase in population purely due to reproduction. $\beta_r$ and $\beta_u$ are the sexual growth rate constants of the rural and urban populations, respectively. The second term of the two equations is due to rural-to-urban migration. As argued above, the rate of increase in urban population due to migration at time $t$ has been assumed to depend on the difference in GDP per capita of the urban and rural regions, $E_r(t)$ and $E_u(t)$, respectively. Notice that GDP per capita itself is a function of time. While in a more precise model the rate of migration at time $t$ should depend on the economic disparity observed prior to $t$, since a person's final physical act of migration occurs after observing the economic disparity, in this simple toy model we ignore such details.

Since we intend to model the trend of the last 50 years, in which the empirical data for most countries indicate that the net flow of population is from rural to urban (Fig. \ref{fig:EmpiricalUrbanFrac}), we thus only consider rural-to-urban migration (and not vice versa). This is indicated by the second term in the equations (notice the sign). The rate of rural-to-urban migration multiplicatively depends upon $N_r$ and the difference in the GDP per capita of urban and rural region, $E_u(t) - E_r(t)$. The parameter $k$ is the proportionality constant (referred to here as the \textit{migration rate constant}) denoting the rate of rural-to-urban migration per unit economic disparity. We call it the \textit{migration rate constant}. While $\beta_r$, $\beta_u,$ and $k$ themselves have been changing over the years, in this simplified framework we assume them to be constants in our model.

Further, we assume that both the regions experience exponential economic development such that the GDP per capita (denoted as $E$) grows exponentially in both the regions:

\begin{subequations}\label{eq:ExpnentialGrowthOfEconomy}
    \begin{align}
        E_u(t) &= E_{0u}e^{\alpha_{u}t}, \label{eq:UrbanEconomy}\\
        E_r(t) &= E_{0r}e^{\alpha_{r}t}, \label{eq:RuralEconomy}
    \end{align}
\end{subequations}

where $E_u(0)$ and $E_r(0)$ are the respective values of $E_u$ and $E_r$ at the initial time ($t=0$), and $\alpha_u$, $\alpha_r$ are the exponents. While the time dependence of GDP per capita has been a subject of intense debate, with claims of linear, exponential, and logistic growth \cite{kwasnicki2013logistic, lange2018mature}, in this simple toy model we assume exponential growth.\\

From $N_u(t)$ and $N_r(t)$ we can define the urban fraction, $f(t)$,  at time $t$ as:

\begin{equation}
 f(t) = \frac{N_u(t)}{N_u(t) + N_r(t)}.
\end{equation}

Using Eq. \eqref{eq:AbsoluteRateEqns} and Eq. \eqref{eq:ExpnentialGrowthOfEconomy} one can obtain the rate of change of the urban fraction (see appendix for details):

\begin{eqnarray}\label{eq:urbanization_equation}
\dot f &=& k(1-f)(E_{0u}e^{\alpha_ut} - E_{0r}e^{\alpha_rt}) -\Delta\beta f(1-f),
\end{eqnarray}
where $\Delta\beta = \beta_r - \beta_u$.\\

We refer the above Eq. \eqref{eq:urbanization_equation} as the \textit{urbanization equation}. Notice that it depends upon various demographic and economic parameters such as $\Delta\beta$ (which is the difference of the rural and urban reproduction rates), $k$ (migration rate constant), the two exponents $\alpha_u$ and $\alpha_r$, and $E_{0u}$ (urban GDP per capita at $t=0$), $E_{0r}$ (rural GDP per capita at $t=0$). Since we want to capture the urbanization dynamics for the last 50 years, i.e., from 1970 to 2020, hence for us $t=0$ corresponds to the year 1970.

\subsection*{Parameter estimation}
Using historical data we first analytically obtained theoretical estimates of our coarse-grained parameters. These values were then used as initial guesses for computational calibration via nonlinear least-squares fitting.

\subsubsection*{Theoretical estimation}
We first estimated the values of the parameters of our model for USA from empirically reported data. Our parameter estimation procedure is coarse with an aim of obtaining ballpark values which can be used to test whether our model qualitatively reproduces the empirically observed urbanization curve. Below we present our estimation technique (highlighted for USA).\\

\textit{(i) Estimation of $E_{0u}$ and $E_{0r}$}: While the year-wise data for GDP per capita is readily available for most countries, our urbanization equation requires the urban and rural GDP per capita separately, which is difficult to trace. But the rural and urban median household (annual) income is a commonly reported statistic in census reports, and since the household income is positively correlated to the GDP per capita \cite{diacon2015relationship}, we hence make use of the following approximation: For a particular year, the ratio of the median urban household income (\textit{UHI}) to that of the median rural household income (\textit{RHI}) can be taken as a proxy for the ratio of the urban and rural GDP per capita for that year. I.e., 

\begin{equation*}
    \frac{UHI (t)}{RHI(t)} \approx \frac{E_u(t)}{E_r(t)}
\end{equation*}

Using the above equation for the year 1970 (when \textit{UHI} = $\$4409$ and \textit{RHI} = $\$3217$ \cite{BEA_Regional_GDP}, we hence obtain

\begin{equation}\label{eq:E-1}
\frac{E_u(1970)}{E_r(1970)} \approx \frac{4409}{3217} \approx 1.37.
\end{equation}

Since we want to estimate $E_u$ and $E_r$, we need another equation relating them. Since we know that (i) the average GDP per capita of USA ($E$) in 1970 was $E\approx \$25232$ and (ii) fraction of population living in rural areas was $f_r \approx 0.26$, we hence can write:

\begin{equation}\label{eq:E-2}
    E = f_r E_r + (1-f_r)E_u,
\end{equation}
for the year 1970.

From Eq. \eqref{eq:E-1} and Eq. \eqref{eq:E-2} we obtain $E_u(1970)=\$27168$ and $E_r(1970) = \$19831$. Notice that the above calculated $E_u(1970)$ and $E_r(1970)$ are also our initial GDP per capita, i.e. $E_{0u}$ and $E_{0r}$, since our data series starts at 1970. Thus $E_{0u} = \$27168$ and $E_{0r}=\$19831$.\\

\textit{(ii) Estimation of $\alpha_{u}$ and $\alpha_{r}$}: To determine $\alpha_{u}$ and $\alpha_{r}$, it is necessary to have more data points for $E_u$ and $E_r$ for different years; the corresponding exponents can then be extracted by an exponential fit. Since we could only locate the median UHI and RHI data for USA for the years 1970, 1980, 1990, 2000, 2010 and 2020, we hence could estimate only six sets of values for $E_u$ and $E_r$ for the above-mentioned years (see appendix for details). The exponents hence obtained were $\alpha_u=0.0168$ and $\alpha_r=0.0173$ (see appendix for details).\\ 

\textit{ (iii) Estimation of $k$}: As done for the other parameters, we make a crude estimation of $k$, the migration rate constant. Eq. \eqref{eq:AbsoluteRateEqns} suggests that the number of people migrating from rural to urban region in a period between $t$ and $t+\Delta t$ is:
\begin{equation*}
    \Delta N = kN_r(t)\biggl(E_u(t) - E_r(t)\biggr)\Delta t.
\end{equation*}

\cite{jones2020urban} reports that $\approx$ 13.2 million people migrated from rural to urban regions in the period 1990-2000 and $\approx$ 13.3 million people migrated in the period 2000-2010. Making a rough estimation of $k$, we use the above equation for the two time periods. I.e., for 1990-2000: using $E_u$, $E_r$ and $N_r$ at $t=1990$ and $\Delta t=10$ and $\Delta N_u = 13.2$x$10^{6}$ we obtain $k\approx 3.3$x$10^{-6}$ per year per dollar. Similarly for 2000-2010: using $E_u$, $E_r$ and $N_r$ at $t=2000$ and $\Delta t=10$ and $\Delta N_u = 13.3$x$10^{6}$ we obtain $k\approx 2.6$x$10^{-6}$ per year per dollar. We take the average value of the two estimates as our final theoretical estimate for $k$, i.e., $2.95$x$10^{-6}$ per year per dollar\\

\textit{ (iv) Estimation of $\Delta \beta$ (=$\beta_r - \beta_u$)}: Data suggests that the difference between rural and urban  \textit{total fertility rate} (average number of children that are born to a woman over her lifetime) has risen from 0.11 in 2007 to 0.24 in 2017 \cite{ely2018trends}. Notice that the reproduction rates ($\beta_r, \beta_u$) in our population rate equations, Eq. \eqref{eq:AbsoluteRateEqns}, differ from the fertility rate. The rate of increase of urban population (due to reproduction) using the fertility rate constant (say $\gamma_u$) would be $\dot{N_u} = \gamma_u N_u^{w}(t)$ (where $\gamma$ is the fertility rate and $N_u^w(t)$ is the number of women in the urban population at time $t$). Whereas in our case the rate of increase of population due to reproduction has been assumed to depend upon the total population, i.e., $\dot{N_u} = \beta_u N_u(t)$ and not only on the number of women $N_u^{w}(t)$. To reconcile the two, we assume that the number of women remains a constant fraction of the total population (say half), i.e., $N_u^w(t) = N_u(t)/2$. Using this in the fertility rate equation we get $\dot{N_u} = \gamma_u N_u(t)/2$. Comparing this equation with our original rate equation we get $\beta_u = \gamma_u/2$. Similarly, $\beta_r = \gamma_r/2$. We thus obtain that $\Delta \beta = \beta_r - \beta_u = (\gamma_r - \gamma_u)/2 = 0.11/2 = 0.055$ in 2007 and $0.24/2 = 0.12$ in 2017. We thus assume the average of the two values, i.e., 0.0875 per year as a crude estimate for $\Delta \beta$.

\subsubsection*{Calibrated estimation}
\begin{figure}[ht]
\centering
\includegraphics[width=14cm,height=7cm]{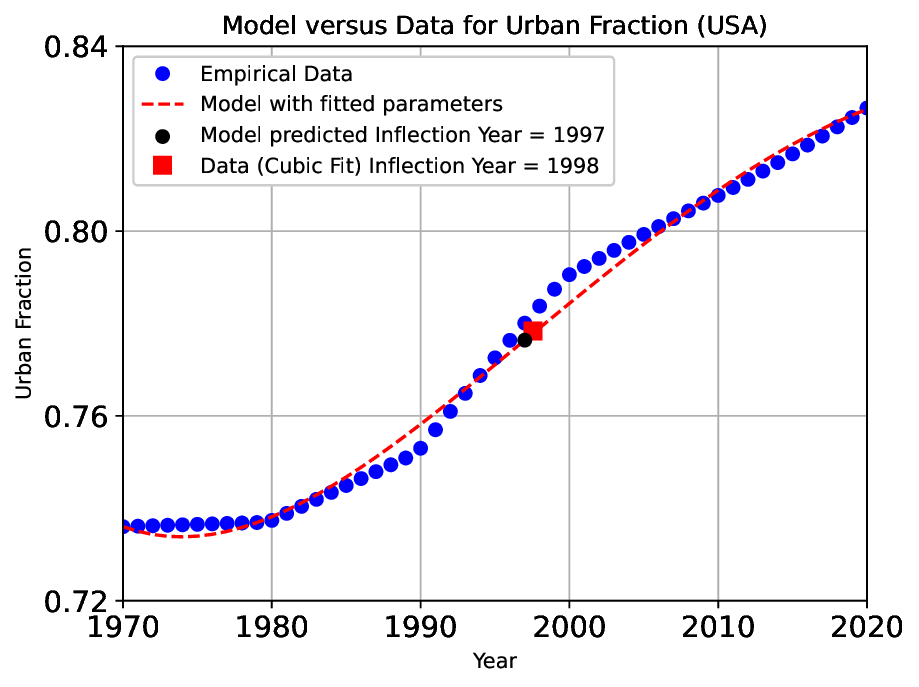}
\caption{\textbf{Model versus empirical data for USA.} Blue dotted curve highlight the empirical data of urban fraction of USA for the last 50 years and red dashed curve is the model-predicted urbanization trajectory obtained by Eq. \eqref{eq:urbanization_equation} using the calibrated values of the parameters.
The figure also highlights the inflection year (when slope changes from positive to negative). Black square (year 1998) represents the inflection year computed by finding the inflection in the cubic polynomial fit to the empirical urban fraction data of USA. Magenta square (year 1999) represents the inflection year for the urbanization trajectory predicted by the model Eq. \eqref{eq:urbanization_equation}.}
\label{fig:UrbanFracSimulationUSA}
\end{figure}

Using the above obtained theoretical estimates as initial guess values, we obtained the best fit values of the parameters for which the urbanization equation Eq. \eqref{eq:urbanization_equation} fits best with the empirical data. Nonlinear least squares optimization method was employed for this purpose. The theoretical and calibrated best-fit estimates of the parameters are reported in Table \ref{ParameterTableUSA}. The trajectory of the model using the best-fit (calibrated) parameter values is shown in Fig. \ref{fig:UrbanFracSimulationUSA}.

\begin{table}[!h]
\begin{center}
\caption{Parameter estimation for USA.\\}
 \vspace{1mm}
\label{ParameterTableUSA}
\begin{tabular}{ |p{3cm}||p{4cm}|p{4cm}|  }
 \hline
 Parameter (units)& Theoretical estimate & Calibrated estimate\\
 \hline
 $E_{0u}$ (\$) & 27168  & 27168\\
 $E_{0r}$ (\$) & 19831  & 19831\\
 $\alpha_u$ (/year) & 0.0168  & 0.0166 \\
 $\alpha_r$ (/year) & 0.0173 & 0.0198\\
 $\Delta\beta$ (/year) &   0.0875 & 0.1806\\
 $k$ (/year /\$) & $2.95$x$10^{-6}$ & $1.80$x$10^{-5}$\\
 \hline
\end{tabular}
\end{center}
\end{table}

Using values in the ballpark of the estimated values mentioned above for the parameters, we simulated the urbanization equation (Eq. \eqref{eq:urbanization_equation}).



\section*{Discussion and Conclusion}
Analysis of the urban fraction data for 16 populous countries for the last 50 years displayed an increasing trend but with four qualitatively different shapes - (i) a continuous accelerating growth, (ii) a continuous decelerating growth, (iii) a two-phase growth starting from acceleration and transitioning to deceleration, and (iv) two-phase growth transitioning from deceleration to acceleration. In this manuscript we have tried to build a simple model to understand these qualitatively different types of urbanization trajectories (curve of urban population fraction with time). 

While rural to urban migration will be influenced by a large variety of factors, we assume that the economic inequality between rural and urban regions is the key determinant of human migration. We hence used the difference in GDP per capita of urban and rural regions to determine the rate of human flow from one region to another. Using a coarse-grained approach, we assume that a country can be divided into two regions, urban and rural. Population of each region changes due to: (i) sexual growth rate, and (ii) migration from rural to urban region. The migration rate depends upon the difference in the GDP per capita of the urban and the rural region. Further, the GDP per capita of both the regions in turn increase exponentially with separate exponents. We show that this simple model with four variables (rural and urban population and GDP per capita) and a few coarse economic and demographic parameters can generate all the four empirically observed urbanization trends. An interesting observation from our model is that it is the difference between the rural and urban sexual growth rates, rather than the absolute values, that affects urbanization dynamics.

Although the demographic parameters cannot be assumed to be constants due to the ever-changing socio-economic landscape of a country, we assume that for a small time frame (50 years) it can be assumed to not change much. With this assumption we estimated the values of the coarse-grained parameters for USA using empirical data for the last 50 years which were then used to obtain the optimized values of the parameters that produce the best-fit urbanization trajectory.

Theoretical exploration of our model can be used to study the characteristics of the inflection point of the urbanization curve. Identifying the inflection point and understanding the factors that influence its timing could be essential to effectively manage and guide urbanization. To control urbanization, while an early inflection is desirable in the 'Acceleration to Deceleration' cases, on the other hand in the 'Deceleration to Acceleration' case one would want to delay the approach to the inflection point. Empirical evidence suggests that the occurrence of the inflection point varies across regions (Fig. \ref{fig:EmpiricalUrbanFrac}). A theoretical study of our model can determine the conditions for inflection to occur and also the mathematical expression for the inflection point and show how the economic and migration parameters influence it. 

A shortcoming of our model is that our coarse-grained toy model
consists only of two regions - urban and rural. However, an actual country could have several urban and rural areas with their own economic and demographic parameters. While realistic, such a situation would complicate the model that would need a network theoretical approach, and estimation of a large number of parameters. A model of such character is under construction by our group.

\bibliographystyle{apalike}
\bibliography{references}

\newpage
\section*{Appendix}
\subsection*{Parameter estimation via nonlinear least squares optimization}
We calibrated the parameters of the urbanization dynamics model by minimizing the squared error between the model output and observed urban fraction data for the USA. This parameter estimation was carried out via nonlinear least squares optimization using the L-BFGS-B method, with numerical integration of the ODE system performed using a stiff solver.

The optimization was performed using the L-BFGS-B algorithm (via \textit{scipy.optimize.minimize} in Python), subject to physically reasonable bounds on the parameters. The ODE system was numerically integrated using a stiff solver (solve\_ivp with the BDF method), and the model fit was evaluated over the same time points as the empirical data.

\end{document}